# THE THEORY OF MULTIPLE PEELING

Nicola M. Pugno

*Dept. of Structural Engineering and Geotechnics, Politecnico di Torino, Corso Duca degli Abruzzi 24, 10129, Torino, ITALY*
*Laboratory of Bio-inspired Nanomechanics Giuseppe Maria Pugno*
*Tel: +39 011 564 4902; Fax: +39 011 564 4899; Mobile: +334 3097397; Email: nicola.pugno@polito.it; Skype: nicola.pugno;*
*Webpage: http://staff.polito.it/nicola.pugno/*

*National Institute of Nuclear Physics (INFN), National Laboratories of Frascati, Via E. Fermi 40, 00044, Frascati, ITALY*

*National Institute of Metrological Research (INRIM), Strada delle Cacce 91, I-10135, Torino, ITALY*

*Consorzio Nazionale Interuniversitario per le Scienze Fisiche della Materia (CNISM), via della Vasca Navale 84, 00146, Roma, ITALY*

**Abstract**

In this paper we solve the multiple peeling problem by applying a fracture mechanics approach to a complex system of films, adhering to the substrate and having a common hinge, where the pulling force is applied. The simplest V-shaped system, consisting of two identical peeling tapes is considered as a case study (to be solved coupling six nonlinear equations); an optimal peeling angle, at which adhesion is maximal, is discovered.

## 1. Introduction

In spite of the interest of the fracture mechanics community on peeling, the Kendall (1975) model remains the universally adopted theory for single peeling. Its extension to multiple peeling has never been formulated and is the aim of the present paper.

## 2. The theory of multiple peeling

Let us consider a three-dimensional complex system composed by *N* adhesive tapes converging to a common point P, where an external force $\overline{F}$ is applied.

Each tape has cross-section area $A_i$, Young modulus $Y_i$, length $l_i$ and orientation defined by the unitary vector $\bar{n}_i$, see Figure 1.

The elastic displacement $\delta\bar{\eta}$ (assumed to be small, i.e. tape orientations do not change significantly) of the point P can be calculated as follows. The elongation of each tape is $\delta l_i = \delta\bar{\eta} \times \bar{n}_i$, thus the tape tension (if negative, the corresponding tape does not work, and the external load is supported by the other tapes) is $\bar{T}_i = \delta l_i \bar{n}_i Y_i A_i / l_i = k_i \delta\bar{\eta} \times \bar{n}_i \bar{n}_i$, where $k_i = Y_i A_i / l_i$ is the tape stiffness. The equilibrium of the material point (hinge) P, where the load is applied, imposes $\sum_{i=1}^{N} \bar{T}_i = \bar{F}$ or equivalently $[K]\delta\bar{\eta} = \bar{F}$, where $[K]$ is the known (by comparing the last two equations) stiffness matrix of the system. The elastic displacement $\delta\bar{\eta}$ is thus calculated as:

$$\delta\bar{\eta} = [K]^{-1} \bar{F} \qquad (1a)$$

from which the tape elongations $\delta l_i$, tensions $T_i$ and strains $\varepsilon_i$ can be evaluated:

$$\delta l_i = \delta\bar{\eta} \times \bar{n}_i, \quad T_i = k_i \delta l_i, \quad \varepsilon_i = \delta l_i / l_i, \quad i=1,\ldots,N \qquad (1b)$$

Imagine to impose a finite (the tape orientations change significantly) displacement $\Delta\bar{\eta}$ at the point P, to be accommodated by multiple *virtual* delaminations $\Delta l_i$ and elastic elongations of the tapes. A new global configuration, denoted by the symbol prime, takes place, see Figure 2.

From the scheme reported in Fig. 2 we deduce the validity of the following equations:

$$\bar{l}_i(1+\varepsilon_i) + \Delta\bar{l}_i + \Delta\bar{\eta} = \bar{l}_i'(1+\varepsilon_i'), \quad \bar{l}_i = l_i \bar{n}_i, \quad \bar{l}_i' = (l_i + \Delta l_i)\bar{n}_i', \quad i=1,\ldots,N \qquad (2)$$

The strains $\varepsilon_i$ are known and their current values $\varepsilon_i'$ can be derived, according to eq. (1), as a function of the unknown orientations $\bar{n}_i'$. Accordingly, coupling eqs. (1b) and (2), we can write $4N$ scalar equations in $4N$ unknowns: the $N$ amplitudes of the virtual delaminations $\Delta l_i$ (their directions are known *a priori* from the configuration of adhering tapes), the $N$ current strains $\varepsilon_i'$ and the $2N$ significant components of the new tape orientations $\bar{n}_i'$ ($n_i' = 1$).

Inverting the previous problem, assuming as known three delamination amplitudes in eq. (2), we could derive the other compatible delaminations as well the displacement $\Delta\bar{\eta}$ of the point P. This means that only three virtual delaminations can be considered as independent.

The virtual forces $F_i$ required for the delamination of the $i^{th}$ tape can be calculated by the Griffith's energy balance. Accordingly, the delamination takes place when:

$$-\partial \Pi / \partial l_i = 2\gamma_i w_i, \quad \Pi = E - W, \quad i=1,\ldots,N \qquad (3a)$$

where $\Pi$ is the total potential energy, $E$ is the elastic energy, $W$ is the external work, $\gamma_i$ is the surface energy of the $i^{th}$ tape/substrate interface and $w_i$ is its width.
The elastic energy variation can be calculated as:

$$\Delta E = \frac{1}{2} \sum_{i=1}^{N} A_i l_i Y_i \left( \varepsilon_i'^2 - \varepsilon_i^2 \right) \qquad (3b)$$

The variation of the external work is:

$$\Delta W = \overline{F} \times \Delta \overline{\eta} \qquad (3c)$$

The real critical force is:

$$F_C = \min\{F_i\} = F_j \qquad (3d)$$

and corresponds to the delamination of the $j^{th}$ tape.
The algebraic system is nonlinear but can be linearized considering the differentials instead of the finite differences (e.g. $\Delta \overline{\eta} \to d\overline{\eta}$). However note that the physical system remains intrinsically geometrically nonlinear due to the existence of the orientation variations. Moreover, the energy balance remains non linear in the force $F$.

**3. Double peeling**

The developed treatment is here applied to study a double peeling system, Figure 3. From eq. (1) we derive:

$$T_1 = F \frac{\sin(\theta + \alpha_2)}{\sin(\alpha_1 + \alpha_2)}, \quad T_2 = F \frac{\sin(\theta - \alpha_1)}{\sin(\alpha_1 + \alpha_2)}, \quad \varepsilon_i = T_i / (Y_i A_i)$$

The previous equations are valid for $T_{1,2} > 0$ thus for $\alpha_1 < \theta < \pi - \alpha_2$. If a tension is negative only the other tape sustains the entire load and thus we have a classical single peeling (if both the tensions are negative the load cannot be in equilibrium). From eq. (2) we have:

$$\begin{cases} l_1(1+\varepsilon_1)\cos\alpha_1 + \Delta l_1 + \Delta u = (l_1 + \Delta l_1)(1+\varepsilon_1')\cos(\alpha_1 + \Delta\alpha_1) \\ l_1(1+\varepsilon_1)\sin\alpha_1 + \Delta v = (l_1 + \Delta l_1)(1+\varepsilon_1')\sin(\alpha_1 + \Delta\alpha_1) \\ l_2(1+\varepsilon_2)\cos\alpha_2 + \Delta l_2 - \Delta u = (l_2 + \Delta l_2)(1+\varepsilon_2')\cos(\alpha_2 + \Delta\alpha_2) \\ l_2(1+\varepsilon_2)\sin\alpha_2 + \Delta v = (l_2 + \Delta l_2)(1+\varepsilon_2')\sin(\alpha_2 + \Delta\alpha_2) \\ \varepsilon_1' = \dfrac{F}{Y_1 A_1} \dfrac{\sin(\theta + \alpha_2 + \Delta\alpha_2)}{\sin(\alpha_1 + \alpha_2 + \Delta\alpha_2 + \Delta\alpha_1)} \\ \varepsilon_2' = \dfrac{F}{Y_2 A_2} \dfrac{\sin(\theta - \alpha_1 - \Delta\alpha_1)}{\sin(\alpha_1 + \alpha_2 + \Delta\alpha_2 + \Delta\alpha_1)} \end{cases}$$

where $\Delta u$ and $\Delta v$ are the horizontal and vertical components of the displacement $\Delta\bar{\eta}$. Note that the classical single peeling only requires one equation, since no angle and strain variations occur during delamination.

Considering $\Delta\varepsilon_i = \varepsilon_i' - \varepsilon_i$ and solving the previous system in the limit of small variations (i.e. substituting the finite differences with the differentials), yields:

$$[A] = \begin{bmatrix} 1 & 0 & l_1(1+\varepsilon_1)\sin\alpha_1 & 0 & -l_1\cos\alpha_1 & 0 \\ 0 & 1 & -l_1(1+\varepsilon_1)\cos\alpha_1 & 0 & -l_1\sin\alpha_1 & 0 \\ -1 & 0 & 0 & l_2(1+\varepsilon_2)\sin\alpha_2 & 0 & -l_2\cos\alpha_2 \\ 0 & 1 & 0 & -l_2(1+\varepsilon_2)\cos\alpha_2 & 0 & -l_2\sin\alpha_2 \\ 0 & 0 & \sin(\theta+\alpha_2)\cos(\alpha_1+\alpha_2) & \sin(\theta-\alpha_1) & \sin^2(\alpha_1+\alpha_2)Y_1 A_1/F & 0 \\ 0 & 0 & \sin(\theta+\alpha_2) & \sin(\theta-\alpha_1)\cos(\alpha_1+\alpha_2) & 0 & \sin^2(\alpha_1+\alpha_2)Y_2 A_2/F \end{bmatrix}$$

$$[b_1] = \begin{bmatrix} (1+\varepsilon_1)\cos\alpha_1 - 1 \\ (1+\varepsilon_1)\sin\alpha_1 \\ 0 \\ 0 \\ 0 \\ 0 \end{bmatrix}, \quad [b_2] = \begin{bmatrix} 0 \\ 0 \\ (1+\varepsilon_2)\cos\alpha_2 - 1 \\ (1+\varepsilon_2)\sin\alpha_2 \\ 0 \\ 0 \end{bmatrix}, \quad [dx] = \begin{bmatrix} du \\ dv \\ d\alpha_1 \\ d\alpha_2 \\ d\varepsilon_1 \\ d\varepsilon_2 \end{bmatrix}$$

$$[dx] = [A]^{-1}([b_1]dl_1 + [b_2]dl_2)$$

Eq. (3b) in the limit of small variations, gives:

$$dE = Y_1 A_1 l_1 \varepsilon_1 d\varepsilon_1 + Y_2 A_2 l_2 \varepsilon_2 d\varepsilon_2 + \frac{1}{2} Y_1 A_1 \varepsilon_1^2 dl_1 + \frac{1}{2} Y_2 A_2 \varepsilon_2^2 dl_2$$

as well as eq. (3c) poses:

$$dW = F\cos\theta du + F\sin\theta dv$$

According to eq. (3a) and (3d) the delamination force can now be easily obtained.

For example, considering the symmetric case ($\alpha_1 = \alpha_2 = \alpha$, $l_1 = l_2 = l$, $\theta = \pi/2$, and consequently $u = 0$ and $\varepsilon_1 = \varepsilon_2 = \varepsilon$) we find the following solutions:

$$d\varepsilon = \frac{1-(1+\varepsilon)\cos\alpha}{\dfrac{1+\varepsilon}{\varepsilon}\dfrac{\sin^2\alpha}{\cos\alpha}+\cos\alpha}\frac{dl}{l} \cong \frac{(1-\cos\alpha)\cos\alpha}{\sin^2\alpha}\varepsilon\frac{dl}{l}$$

$$d\alpha = \frac{[(1+\varepsilon)\cos\alpha - 1]dl + l\cos\alpha\, d\varepsilon}{l(1+\varepsilon)\sin\alpha} \cong \frac{\cos\alpha - 1 + \varepsilon + l\cos\alpha\, d\varepsilon/dl}{\sin\alpha}\frac{dl}{l}$$

$$dv = l(1+\varepsilon)\cos\alpha\, d\alpha + l\sin\alpha\, d\varepsilon + (1+\varepsilon)\sin\alpha\, dl$$

The previous equations have been linearized in $\varepsilon$. Accordingly, the energy balance is self-consistently written considering terms up to the second power of $\varepsilon$. The result yields:

$$\varepsilon^2 + 2(1-\cos\alpha)\varepsilon - 4\lambda = 0, \quad \text{where } \lambda = \frac{\gamma}{tY}$$

and $t = A/w$ is the tape thickness. Solving this equation, the critical value of the strain $\varepsilon_C$ for delamination is obtained. The previous equation is surprising: it is identical to that of the single peeling problem. However the force required for delamination is different since here we have:

$$F_C = 2YA\varepsilon_C \sin\alpha$$

thus only for $\alpha = \pi/2$ the prediction is that of the single peeling tape loaded by a force $F/2$, for which $F_C = 2YA\varepsilon_C$, as it must physically be.

The delamination force is thus:

$$F_C = 2YA\sin\alpha\left(\cos\alpha - 1 + \sqrt{(1-\cos\alpha)^2 + 4\lambda}\right)$$

The behaviour is depicted in Figure 4. An angle for optimal adhesion $\alpha_{opt}$ clearly emerges as a function of the parameter $\lambda$.

## 3. Conclusion

Herewith we have solved the multiple peeling problem. The system consisting of two peeling tapes has been considered as a case study. For such a case, we have surprisingly observed (i) a governing equation for the strain identical to that of the single peeling and (ii) an optimal peeling angle, at which adhesion is maximal. The last result is of a great importance for the explanation of the functional mechanism of biological adhesives and for adhesive technology as well.

## References

Kendall, K. 1975, Thin-film peeling-the elastic term. J. Phys. D: Appl. Phys. 8, 1449-1452.

**FIGURES**

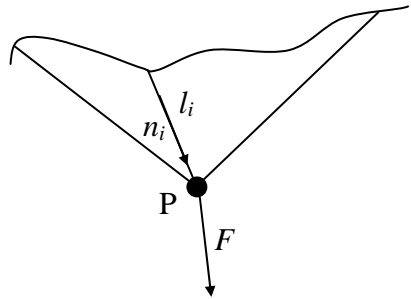

Figure 1. Diagram of the multiple peeling system considered in this study.

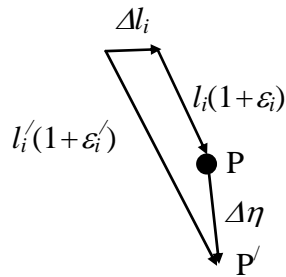

Figure 2. Finite delamination of the *i*th tape.

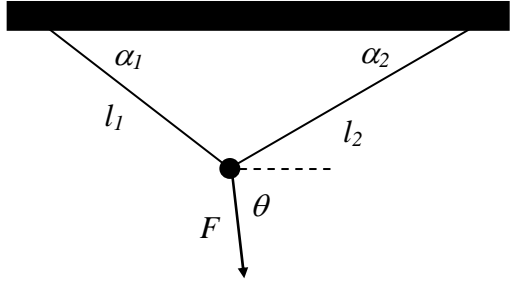

Figure 3. Diagram of the double peeling system considered in this study.

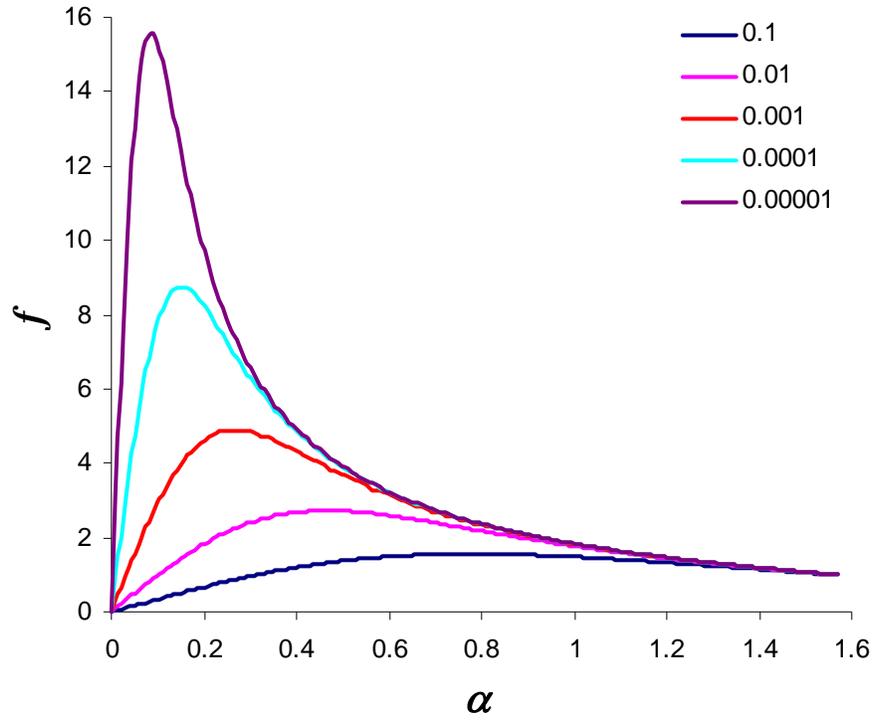

Figure 4. Dimensionless force $f = F_C(\alpha)/F_C(\alpha = \pi/2)$ versus angle $\alpha$ by varying the dimensionless adhesion strength $\lambda$; $F_C(\alpha = \pi/2) = 2YA\left(-1 + \sqrt{1 + 4\lambda}\right)$.